\documentclass[prd,aps,twocolumn,amsmath,amssymb,nofootinbib,preprintnumbers]
{revtex4}

\usepackage{graphicx}
\usepackage{dcolumn}
\usepackage{bm}
\usepackage{amsmath}
\usepackage{amsthm}
\usepackage{amsfonts}
\usepackage{euscript,bbm}
\usepackage{ifthen}
\usepackage{psfrag}
\usepackage{slashed}

\def\ls{\mathrel{\lower4pt\vbox{\lineskip=0pt\baselineskip=0pt
           \hbox{$<$}\hbox{$\sim$}}}}
\def\gs{\mathrel{\lower4pt\vbox{\lineskip=0pt\baselineskip=0pt
           \hbox{$>$}\hbox{$\sim$}}}}
\def\drawbox#1#2{\hrule height#2pt
\hbox{\vrule width#2pt height#1pt \kern#1pt
              \vrule width#2pt}
              \hrule height#2pt}

\def\Asym#1#2{\vcenter{\vbox{\drawbox{#1}{#2}
              \kern-#2pt       
              \drawbox{#1}{#2}}}}

\def\nn{\nonumber}

\newcommand{\be}{\begin{equation}}
\newcommand{\ee}{\end{equation}}
\newcommand{\bea}{\begin{eqnarray}}
\newcommand{\eea}{\end{eqnarray}}

\newcommand{\gsim}{\lower.7ex\hbox{$\;\stackrel{\textstyle>}{\sim}\;$}}
\newcommand{\lsim}{\lower.7ex\hbox{$\;\stackrel{\textstyle<}{\sim}\;$}}

\newcommand{\vo}{\mathcal{V}}

\newcommand{\ben}{\begin{enumerate}}
\newcommand{\een}{\end{enumerate}}
\newcommand{\bei}{\begin{itemize}}
\newcommand{\eei}{\end{itemize}}
\newcommand{\bi}{\begin{itemize}}
\newcommand{\ei}{\end{itemize}}
\newcommand{\mc}{\mathcal}

\begin{document}

\title{Correlation between Dark Matter and Dark Radiation in String Compactifications}

\author{Rouzbeh Allahverdi$^{1}$}
\author{Michele Cicoli$^{2,3,4}$}
\author{Bhaskar Dutta$^{5}$}
\author{Kuver Sinha$^{6}$}

\affiliation{$^{1}$~Department of Physics and Astronomy, University of New Mexico, Albuquerque, NM 87131, USA \\
$^{2}$~Dipartimento di Fisica ed Astronomia, Universit\`a di Bologna, via Irnerio 46, 40126 Bologna, Italy \\
$^{3}$~INFN, Sezione di Bologna, 40126 Bologna, Italy \\
$^{4}$~Abdus Salam ICTP, Strada Costiera 11, Trieste 34014, Italy \\
$^{5}$~Mitchell Institute for Fundamental Physics and Astronomy, Department of Physics and Astronomy,
Texas A\&M University, College Station, TX 77843-4242, USA \\
$^{6}$~Department of Physics, Syracuse University, Syracuse, NY 13244, USA
}

\begin{abstract}
Reheating in string compactifications is generically driven by the decay of the lightest modulus
which produces Standard Model particles, dark matter and light hidden sector degrees of freedom
that behave as dark radiation. This common origin allows us to find an interesting correlation between dark matter and dark radiation.
By combining present upper bounds on the effective number of neutrino species $N_{\rm eff}$ with
lower bounds on the reheating temperature as a function of the dark matter mass $m_{\rm DM}$ from Fermi data,
we obtain strong constraints on the $(N_{\rm eff},m_{\rm DM})$-plane.
Most of the allowed region in this plane corresponds to non-thermal scenarios
with Higgsino-like dark matter. Thermal dark matter can be allowed only if $N_{\rm eff}$ tends to its Standard Model value.
We show that the above situation is realised in models with perturbative moduli stabilisation
where the production of dark radiation is unavoidable since bulk closed string axions remain light and do not get eaten up by anomalous $U(1)$s.
\end{abstract}
MIFPA-14-02
\maketitle

\section{Introduction}

The cosmological history of dark matter (DM) -- whether it is thermal or non-thermal --
is at present under a lot of theoretical and experimental investigation.
In large parts of the parameter space of most particle physics models,
DM is either thermally under- or overproduced. On the experimental front,
Fermi data have already ruled out some range of masses for thermal DM in particular annihilation modes \cite{fermi}.
LHC constraints have also started to restrict the parameter space of some supersymmetric models to underabundance scenarios.
It is possible that upcoming LHC data and various direct and indirect DM detection experiments
will illuminate the nature of DM in the next few years.

A non-standard cosmological history where DM is produced non-thermally
from the decay of a heavy scalar is well motivated for a number of reasons \cite{Moroi:1999zb}.
A non-thermal history can accommodate both thermally underproduced (Higgsino-like)
as well as thermally overproduced (Bino-like) candidates,
opening up vast regions of the parameter space for supersymmetric models.

Moreover, a non-thermal history arises naturally in string compactifications
since they are characterised by the ubiquitous presence of moduli which
parameterise the size and shape of the extra dimensions \cite{Bobby,DMinSeqLVS}.
Even if in general these moduli $\phi$ acquire large masses of order the gravitino mass,
they are long-lived since they have only Planck-suppressed interactions to ordinary particles.
Hence they have enough time to come to dominate the energy density of the universe before decaying.
Therefore in string compactifications the reheating process occurs due to the decay of the lightest modulus
which produces non-thermally both Standard Model (SM) degrees of freedom and DM particles,
diluting everything that has been produced before.
However since the moduli are gauge singlets, they could in principle decay also to
light hidden sector degrees of freedom which behave as dark radiation (DR).
Two main examples of this kind of particles are light axions \cite{DRinLVS,WinoDMDR} and hidden photons \cite{Cicoli:2011yh}.

This raises the following questions: $(i)$ how generic is the presence of DR in string compactifications?,
and $(ii)$ given that both DM and DR have a common origin from the decay of a string modulus,
is it possible to find a correlation between them by combining DM observational constraints
with present bounds on the effective number of neutrino species $N_{\rm eff}$?

The purpose of this paper is to address these two questions. With regard to the first,
we show that \emph{in string compactifications with perturbative moduli stabilisation, the production of dark radiation is unavoidable}.
In fact, the moduli are complex fields whose real and imaginary parts are, respectively, $\phi$ and an axion-like particle $a$
enjoying a shift symmetry that is preserved at the perturbative level and broken only by non-perturbative effects.
Hence if the moduli develop a non-perturbative potential, both $\phi$ and $a$ get a mass of the same order of magnitude.
In this case, the production of $a$ from the decay of $\phi$ is kinematically forbidden. On the other hand,
if the moduli are stabilised at perturbative level, $\phi$ is heavier than $a$ since the axions remain massless
at this level of approximation because of their shift symmetry \cite{Nflation,Cicoli:2012sz}.
Subleading non-perturbative effects will then slightly lift the
axionic directions. Therefore, in this case $\phi$ would decay to light axions $a$ which behave as DR.
These light axions could still be removed from the low-energy spectrum by getting eaten up by
anomalous $U(1)$ gauge bosons due to the anomaly cancellation mechanism. However, we shall
show that this is never the case for light closed string axions living in the bulk of the extra dimensions.

With regard to the second question, in the presence of DR,
the reheating temperature $T_{\rm rh}$ can be written
as a function of $N_{\rm eff}$ and $m_\phi$.
Thus, the present upper bound on $N_{\rm eff}$ from Planck+WMAP9+ACT+SPT+BAO+HST \cite{Planck}
translates into a lower bound on $T_{\rm rh}$ as a function of $m_\phi$.
Moreover, an independent lower bound on $T_{\rm rh}$ as a function of the DM mass $m_{\rm DM}$
can be obtained from the Fermi satellite which constrains the DM annihilation cross section using dwarf galaxy data \cite{fermi}.
Combining these two results, we find an interesting correlation between DM and DR,
ruling out different regions of the $(N_{\rm eff},m_{\rm DM})$-plane as a function of $m_\phi$ which, in turn,
in supersymmetry breaking scenarios based on gravity mediation, is tied to the soft masses $M_{\rm soft}$.

The allowed regions of parameter space largely prefer non-thermal scenarios with $T_{\rm rh}\gtrsim \mc{O}(1)$ GeV for
values of $m_\phi$ which generically lead to TeV-scale soft terms.
Given that $T_{\rm rh}$ cannot be lowered to values close to $\mc{O}(1)$ MeV,
DM particles would be overproduced by the decay of $\phi$ if subsequently they do not annihilate efficiently.
Thus non-thermal models with Bino-like DM are disfavoured with respect to scenarios where DM is a Higgsino-like LSP.
If Higgsino DM is underproduced, the remaining DM abundance can come from the QCD axion
which can be realised as the phase of an open string mode charged under an anomalous $U(1)$ \cite{WinoDMDR,openAxion}.
Moreover, as the soft masses get heavier, more allowed regions of the $(N_{\rm eff},m_{\rm DM})$-plane open up for smaller DM masses.
Finally, thermal DM can be allowed only if $N_{\rm eff}$ approaches its SM value.

This paper is organised as follows. In Section \ref{Sec1} we review the production of DM and DR from moduli decays,
while in Section \ref{Sec2} we highlight the correlation between DM and DR by establishing the allowed regions on the ($N_{\rm eff},m_{\rm DM}$)-plane
using present bounds on $N_{\rm eff}$ and Fermi data from dwarf spheroidal galaxies.
In Section \ref{Sec3} we show the explicit example of sequestered models within the framework of type IIB LARGE Volume Scenarios (LVS) \cite{sequestering}.
After presenting our conclusions in Section \ref{Concl}, in Appendices \ref{App0} and \ref{App} we analyse in detail
the behaviour of both closed and open string axions.

\section{Dark matter and dark radiation from moduli decays}
\label{Sec1}

\subsection{Reheating temperature}

The lightest modulus $\phi$ can decay to both visible and hidden degrees of freedom with partial decay widths
given, respectively, by $\Gamma_{\rm vis} = c_{\rm vis} \Gamma_0$ and $\Gamma_{\rm hid} = c_{\rm hid} \Gamma_0$
where $c_{\rm vis}$ and $c_{\rm hid}$ are model-dependent prefactors and:
\be
\Gamma_0 = \frac{1}{48\pi} \frac{m_\phi^3}{M_P^2}\,. \nn
\ee
Hence the total decay rate is $\Gamma_{\rm tot} = c_{\rm tot} \Gamma_0$
with $c_{\rm tot} = c_{\rm vis} + c_{\rm hid}$.
If the hidden sector is composed of light axion-like particles with only gravitational couplings, these relativistic decay products
would not thermalise. On the other hand, the modulus decay reheats the visible sector to a temperature
$T_{\rm rh}= \left[30/(\pi^2 g_*)\right]^{1/4} \rho_{\rm vis}^{1/4}$,
where $g_*$ is the number of relativistic degrees of freedom at $T_{\rm rh}$, and
the energy density of the visible sector thermal bath is:
\be
\rho_{\rm vis} = \frac{c_{\rm vis}}{c_{\rm tot}} \rho_{\rm tot} = \frac{3 c_{\rm vis}}{c_{\rm tot}} H^2 M_P^2\,.
\nn
\ee
Using the fact that at $\phi$ decay $3H^2 \simeq 4\Gamma_{\rm tot}^2/3$, one finds:
\be
T_{\rm rh} \simeq \frac{1}{\pi}\left(\frac{5\,c_{\rm vis}c_{\rm tot}}{288 g_*}\right)^{1/4}m_\phi\sqrt{\frac{m_\phi}{M_P}}\,.
\label{Trh1}
\ee

\subsection{Dark radiation production}

In string scenarios where some moduli are fixed perturbatively,
the corresponding axions remain light because of the Peccei-Quinn shift symmetry.
Let us point out that fixing all the moduli via non-perturbative effects seems to be
a rather non-generic situation because of the difficulty to satisfy the instanton zero-mode condition
and the interplay between chirality, the cancellation of Freed-Witten anomalies and non-perturbative effects \cite{CMV}.
Moreover, as we shall show in Appendix \ref{App0}, bulk closed string axions do not get eaten up by anomalous $U(1)$s,
and so they cannot be removed from the low-energy spectrum. These considerations imply that generically $c_{\rm hid}\neq 0$,
and so some DR is produced by the lightest modulus decay \cite{DRinLVS,WinoDMDR}.

Writing $N_{\rm eff}$ as $N_{\rm eff}=N_{\rm eff,SM}+\Delta N_{\rm eff}$,
where the SM value is $N_{\rm eff,SM}=3.04$, we obtain:
\be
\Delta N_{\rm eff} = \frac{43}{7} \frac{c_{\rm hid}}{c_{\rm vis}}\,.
\label{DNeff}
\ee
The present observational bound on $\Delta N_{\rm eff}$ from Planck+WMAP9+ACT+SPT+BAO+HST at $2\sigma$ is
$\Delta N_{\rm eff} = 0.48^{+0.48}_{-0.45}$ \cite{fermi}, implying $\Delta N_{\rm eff}\lesssim 0.96$ at $2\sigma$.
Combining this with (\ref{DNeff}) we find a model-independent constraint between visible and hidden couplings:
\be
c_{\rm vis} \gtrsim  6.4 \,c_{\rm hid}\,.
\label{cvisBound}
\ee

\subsection{Non-thermal dark matter}

DM particles are produced non-thermally from the decay of $\phi$ if $T_{\rm rh}$
is smaller than the freeze-out temperature $T_{\rm f} \simeq m_{\rm DM}/20$ where $m_{\rm DM}$
is the DM mass. In order to obtain TeV-scale soft terms, the mass of $\phi$ in string compactifications is generically of order
$m_\phi\simeq 5\cdot 10^6$ GeV, giving from (\ref{Trh1}), $T_{\rm rh}\simeq \mc{O}(1)$ GeV if $g_*=68.5$
and $c_{\rm vis}\simeq \mc{O}(10)$, as needed to satisfy (\ref{cvisBound}) for $c_{\rm hid}\simeq \mc{O}(1)$.
Hence for $m_{\rm DM}\gtrsim \mc{O}(20)$ GeV, DM particles cannot thermalise after their non-thermal production from $\phi$ decay
which leads to an abundance of DM particles \cite{DMinSeqLVS}:
\bea
\label{DMdens}
{n_{\rm DM} \over s} = {\rm min} \left[\left({n_{\rm DM} \over s}\right)_{\rm obs}
{\langle \sigma_{\rm ann} v \rangle_{\rm f}^{\rm th} \over \langle \sigma_{\rm ann} v \rangle_{\rm f}} \left({T_{\rm f} \over T_{\rm rh}}\right),
Y_{\phi}~ {\rm Br}_{\rm DM} \right]
\eea
where $\langle \sigma_{\rm ann} v \rangle_{\rm f}^{\rm th}\simeq 3 \times 10^{-26} {\rm cm}^3\,{\rm s}^{-1}$
is the value needed in the thermal case to match the observed DM abundance:
\be
\label{obs}
\left({n_{\rm DM} \over s}\right)_{\rm obs} \simeq 5 \cdot 10^{-10} ~ \left({1 ~ {\rm GeV} \over m_{\rm DM}}\right) \,,
\ee
$Y_\phi = 3 T_{\rm rh} / (4 m_\phi)$ is the yield of particle abundance form $\phi$ decay,
and ${\rm Br}_{\rm DM}$ denotes the branching ratio for $\phi$ decays into $R$-parity odd particles
(which subsequently decay to DM) with ${\rm Br}_{\rm DM} \gsim 10^{-3}$.

The first option on the right-hand side of eq.~(\ref{DMdens}) corresponds to the \emph{Annihilation scenario} where DM particles
undergo some annihilation after their production from the modulus decay. This happens if $\langle \sigma_{\rm ann} v \rangle_{\rm f} = \langle \sigma_{\rm ann} v \rangle_{\rm f}^{\rm th}\, (T_{\rm f}/T_{\rm rh})$. Since $T_{\rm rh} < T_{\rm f}$, this scenario can yield the correct DM abundance only if
$\langle \sigma_{\rm ann} v \rangle_{\rm f} > \langle \sigma_{\rm ann} v \rangle_{\rm f}^{\rm th}$ as in cases with thermal underproduction.
This happens for sub-TeV Higgsino-like DM which is motivated in scenarios such as natural SUSY~\cite{naturalsusy}. Wino-like DM is also thermally underproduced for masses up to $\sim 3$ TeV. However, pure Wino DM has a significantly larger annihilation cross section than Higgsino DM. As a result, the bounds set by Fermi data rule out sub-TeV Wino as the dominant component of DM~\cite{winodm}. Furthermore, pure gravity mediation models with negligible anomaly mediated contributions (as those presented in Section \ref{Sec3}) are characterised by unified gaugino masses at the string scale which do not allow a Wino-like LSP.

The second option on the right-hand side of eq.~(\ref{DMdens}) is the \emph{Branching scenario} where DM particles do not undergo further annihilation.
This case can accommodate thermally overproduced Bino-like DM with $\langle \sigma_{\rm ann} v \rangle_{\rm f} < \langle \sigma_{\rm ann} v \rangle_{\rm f}^{\rm th}$ if $Y_\phi$ is appropriately reduced by lowering $T_{\rm rh}$.
This scenario, however, typically requires a very low reheating temperature: ${\cal O}({\rm MeV}) \lesssim T_{\rm rh} \ll {\cal O}({\rm GeV})$.

\section{Correlation between dark matter and dark radiation}
\label{Sec2}

\subsection{Higgsino-like DM from DR constraints}

Given that in perturbatively stabilised string models $c_{\rm hid}\neq 0$, we can write
the reheating temperature $T_{\rm rh}$ as a function of $\Delta N_{\rm eff}$.
Using (\ref{DNeff}), (\ref{Trh1}) becomes:
\be
T_{\rm rh}\simeq \kappa\sqrt{\frac{c_{\rm hid}}{\Delta N_{\rm eff}}}\left(\frac{68.5}{g_*}\right)^{1/4}
\left(\frac{m_\phi}{5\cdot 10^6\,{\rm GeV}}\right)^{3/2}\,0.72\,{\rm GeV}\,,
\label{Trh2}
\ee
where $\kappa\equiv \left(1+\frac{7}{43}\Delta N_{\rm eff}\right)^{1/4}\simeq 1$
for $0.03\lesssim \Delta N_{\rm eff}\lesssim 0.96$ at $2\sigma$ level.
The expression (\ref{Trh2}) allows us to find a connection between DM and DR.
In fact, setting $c_{\rm hid}=1$, $g_*=68.5$ and $m_\phi= 5\cdot 10^6$ GeV,
the observational upper bound $\Delta N_{\rm eff} \lesssim 0.96$
translates into a lower bound on the reheating temperature: $T_{\rm rh}\gtrsim 0.73$ GeV.
Hence we see that DR constraints shed light on what type of LSP particle
can behave as DM. In fact, with this lower bound on $T_{\rm rh}$ the branching scenario
is ruled out and Bino-like DM is disfavoured.
Thus \emph{DR constraints prefer thermally underproduced Higgsino-like DM}.
As we have already pointed out, the correct Higgsino DM abundance can be easily
obtained in the annihilation scenario. On the other hand,
in the thermal case, which corresponds to the limit $\Delta N_{\rm eff}\to 0$ obtained for $c_{\rm vis}\to \infty$,
the remaining DM abundance can come from the QCD axion \cite{Baer:2011hx} (similarly for the annihilation scenario, if needed).

\subsection{Lower bound on DM mass from Fermi data}

\begin{figure}[!htp]
\centering
\includegraphics[width=3.0in]{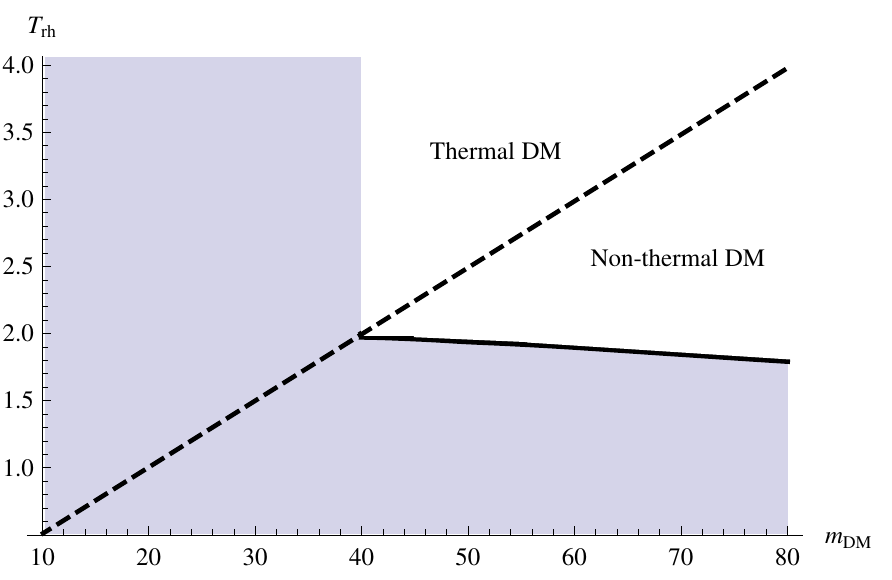}
\caption{Lower bound on $T_{\rm rh}$ (solid line) based on Fermi data for $b \overline{b}$ final state \cite{fermi}.
The result for $WW$ final state is similar. We have taken $T_{\rm f}\simeq \frac{m_{\rm DM}}{20}$ (dashed line). The shaded region is ruled out due to DM overproduction both in the thermal case (for $m_{\rm DM}\lesssim 40$ GeV and above the dashed line) and in the branching scenario
(below the solid and dashed lines).}
\label{Fig1}
\end{figure}

Let us focus on the only allowed case of Higgsino-like DM. Both the standard thermal case and the non-thermal annihilation scenario depend
crucially on the DM annihilation cross section. This is observationally constrained by Fermi data from dwarf spheroidal galaxies
which set an upper bound on the DM annihilation cross section for particular annihilation modes and different DM masses \cite{fermi}.
Using the right-hand side of eq.~(\ref{DMdens}), this upper bound translates into a lower bound on $T_{\rm rh}$
as a function of $m_{\rm DM}$. For simplicity, we can consider that DM particles undergo $S$-wave annihilation which means that the rate of annihilation at the time of freeze-out is approximately the same as that at present time.
This assumption works well in the MSSM when the DM mass is larger than the $W$ mass or close to the bottom quark mass.
The annihilation at present time is constrained by the gamma ray flux from Fermi data~\cite{fermi}. In this paper, we will mainly be taking the dwarf galaxy constraint into account; incorporating results from the galactic center \cite{Center} will not change the main point of our paper.
In Fig. \ref{Fig1} we show the lower bound on $T_{\rm rh}$ based on Fermi data for $b \overline{b}$ final state and $m_{\rm DM}\lesssim 80$ GeV. For $m_{\rm DM}\gtrsim 80$ GeV, the Fermi data can be approximated very well by:
\be
T_{\rm rh}\gtrsim 18\,{\rm GeV}\sqrt{\frac{1\,{\rm GeV}}{m_{\rm DM}}}\qquad\text{for}\quad m_{\rm DM}\gtrsim 80\,{\rm GeV}\,.
\label{Tbound}
\ee

\begin{figure}[!htp]
\centering
\includegraphics[width=3.0in]{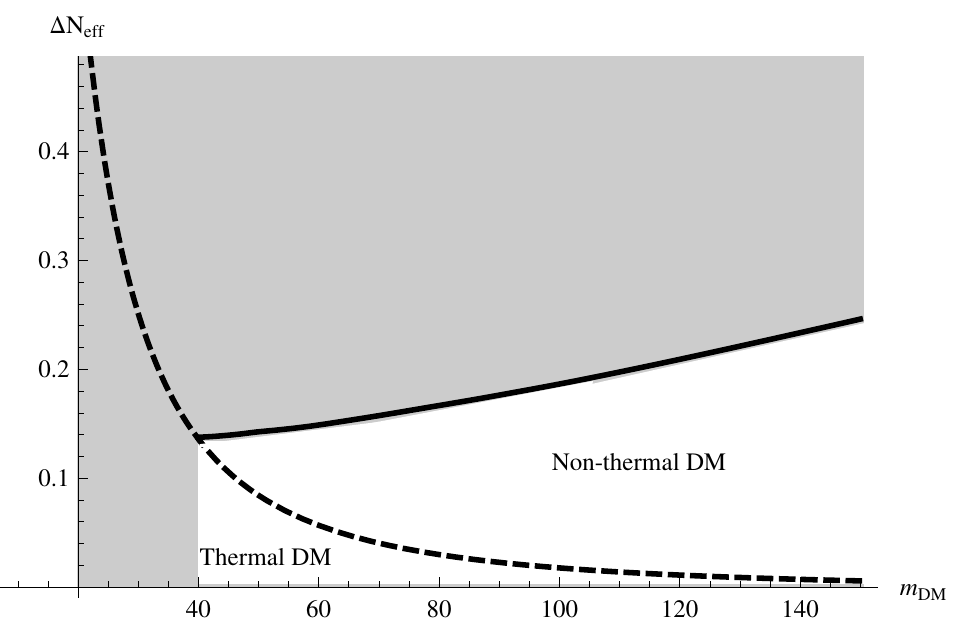}
\caption{Constraints on the $(\Delta N_{\rm eff},m_{\rm DM})$-plane for $c_{\rm hid}=1$,
$g_*=68.5$ and $m_\phi= 5\cdot 10^6$ GeV: the solid line is based on Fermi data
whereas the dashed line represents the freeze-out temperature. The shaded region is ruled out due to DM overproduction
both in the thermal case (for $m_{\rm DM}\lesssim 40$ GeV and below the dashed line) and in the non-thermal branching scenario
(above the solid and dashed lines).}
\label{Fig2}
\end{figure}

Using (\ref{Trh2}), this lower bound on $T_{\rm rh}$ coming from Fermi data can be expressed as an \textit{upper} bound on $\Delta N_{\rm eff}$
as a function of the modulus mass $m_\phi$. Moreover, as we have already seen, an independent \textit{upper} bound on $\Delta N_{\rm eff}$ comes from
cosmological observations. Putting these constraints together in the $(\Delta N_{\rm eff},m_{\rm DM})$-plane (see Fig. \ref{Fig2}),
we find interesting correlations between DM and DR given that only some regions of this plane are allowed.

An important observation that is immediately clear from Fig. \ref{Fig2} is that standard thermal DM scenarios are allowed only for $\Delta N_{\rm eff}\lesssim 0.14$ while the central values from Planck+WMAP9+ACT+SPT+BAO+HST is around $\Delta N_{\rm eff}\simeq 0.48$ \cite{Planck}.\footnote{If also data from X-ray clusters are included, the estimated value for $\Delta N_{\rm eff}$ raises to $\Delta N_{\rm eff}= 0.664\pm 0.29$ at $1\sigma$ \cite{Hu:2014qma}.}
Hence we conclude that \emph{DR constraints combined with Fermi data prefer non-thermal Higgsino-like DM}.
On the other hand, DM can have a thermal history only if
$\Delta N_{\rm eff}$ tends to its SM value. This corresponds to a case where $T_{\rm rh}$
becomes very large because of large values of $c_{\rm vis}$.

\begin{figure}[!htp]
\centering
\includegraphics[width=3.2in]{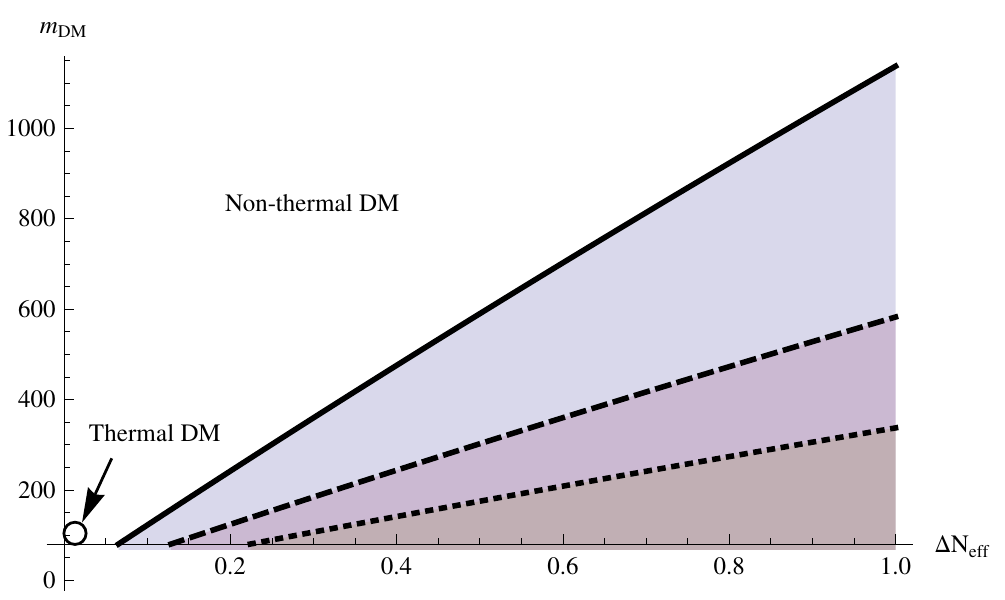}
\caption{Lower bound on the DM mass as a function of $\Delta N_{\rm eff}$ for different values of the modulus mass:
$m_\phi=4\cdot 10^6$ GeV (solid line), $m_\phi= 5\cdot 10^6$ GeV (dashed line) and $m_\phi= 6\cdot 10^6$ GeV (dotted line).
The shaded region is ruled out based on Fermi data \cite{fermi}.
Here we have set $g_*=68.5$ and $c_{\rm hid}=1$.}
\label{Fig3}
\end{figure}

Moreover, an increase in the modulus mass leads to a larger available region for low DM masses.
In fact, the Fermi bound (\ref{Tbound}) can be combined with the expression (\ref{Trh2}) for $T_{\rm rh}$ as a function
of $\Delta N_{\rm eff}$ and $m_\phi$ to give:
\be
m_{\rm DM} \gtrsim \frac{\Delta N_{\rm eff}}{c_{\rm hid}}
\sqrt{\frac{g_*}{68.5}} \left(\frac{5 \cdot 10^5 \,{\rm GeV}}{m_\phi}\right)^3 \,\frac{625\,{\rm GeV}}{\kappa^2}\,.
\label{B2}
\ee
We see that $\Delta N_{\rm eff}$ and $m_\phi$ set a lower bound for the DM mass. The right-hand side of
the inequality (\ref{B2}) is a straight line in $\Delta N_{\rm eff}$ whose slope is set by $m_\phi^{-3}$,
showing how this lower bound becomes weaker from larger values of $m_\phi$ that in string compactifications
correspond to heavier soft terms. This trend is plotted in Fig. \ref{Fig3}.

In the next Section, we investigate these correlations in the context of type IIB sequestered LVS models \cite{sequestering}
where the axion associated to the volume of the extra dimensions remains light and behaves ad DR \cite{DRinLVS,WinoDMDR}
since the corresponding modulus is stabilised perturbatively.

\section{DM-DR correlation in sequestered LVS models}
\label{Sec3}

\subsection{Mass spectrum}

In type IIB LVS models, the lightest modulus $\phi$ parameterises the overall size of the extra dimensions,
and it acquires a mass via perturbative effects of order \cite{LVS}:
\be
m_\phi \simeq m_{3/2}\sqrt{\frac{m_{3/2}}{M_P}}< m_{3/2}\,,
\ee
while the corresponding axion $a$ is almost massless. Moreover, as we shall show in Appendix \ref{App},
this axion is never eaten up by an anomalous $U(1)$, and so its presence and lightness are two generic predictions of LVS models.
Given that $m_\phi<m_{3/2}$, no gravitino problem is induced when the universe is reheated by the decay of $\phi$.
In sequestered LVS models where the visible sector is localised on branes at singularities, the unified gaugino masses at the string scale
are smaller than $m_\phi$ \cite{sequestering}:
\be
M_{1/2}\simeq m_\phi\left(\frac{m_\phi}{M_P}\right)^{1/3} < m_\phi\,,
\label{Mgaugino}
\ee
guaranteeing the absence of any cosmological moduli problem for $M_{1/2}\simeq \mc{O}(1)$ TeV
corresponding to $m_\phi\simeq\mc{O}(5\cdot 10^6)$ GeV and $m_{3/2} \simeq \mc{O}(5\cdot 10^{10})$ GeV.
Depending on the details of the model, the unified scalar masses $m_0$ can instead be either of order $M_{1/2}$
(leading to a mSUGRA/CMSSM~\cite{sugra1} scenario) or of order $m_\phi$ (leading to a split SUSY scenario \cite{split}).

\subsection{Leading decay rates of $\phi$}

When $\phi$ decays, it produces SM degrees of freedom as well as light axions $a$ which behave as DR and contribute to $\Delta N_{\rm eff}$.
The leading decay channels for $\phi$ are to Higgses and
light volume axions $a$ \cite{DRinLVS,WinoDMDR}:

$(i)$ \emph{Decays to Higgses}: the decays to Higgs particles, $\phi \to H_u H_d$, are induced by the Giudice-Masiero term in the K\"ahler potential:
\be \label{KHiggs}
K \supset Z\,\frac{H_u H_d}{2\tau_b}\,,
\ee
where $Z$ is an $\mc{O}(1)$ parameter and $\tau_b$ is the canonically unnormalised field $\phi$ (see Appendix \ref{App}).
The corresponding decay rate is:\footnote{In extensions of the MSSM, one can have a coupling similar to (\ref{KHiggs})
where the Higgses are replaced by new colored fields $X$ and $\overline{X}$ which transform as a $3$ and a $\overline{3}$ under $SU(3)$,
and in `cladogenesis' models play a crucial r\^ole for baryogenesis \cite{Clado}.}
\be
\label{Z}
\Gamma_{\phi \to H_u H_d} = \frac{2 Z^2}{48 \pi} \frac{m_{\phi}^3}{M_P^2}\quad\Rightarrow\quad c_{\rm vis}=2 Z^2\,.
\ee

$(ii)$ \emph{Decays to volume axions}: the axionic partner $a$ of $\phi$ is almost massless, and
so $\phi$ can decay into this particle with decay width:
\be \label{toaxions}
\Gamma_{\phi \to a a} = \frac{1}{48 \pi} \frac{m_{\phi}^3}{M_P^2}\quad\Rightarrow\quad c_{\rm hid}=1\,.
\ee
The decay to other axions, gauge bosons, visible sector scalars and fermions that couple
through non-Giudice-Masiero couplings is suppressed (see \cite{DRinLVS,WinoDMDR} for details).

\subsection{DM-DR correlation}

The first consequence of the present observational upper bound on $\Delta N_{\rm eff}$ is that
the Giudice-Masiero term (\ref{KHiggs}) has to be present with $Z \gtrsim 1.8$ in order to satisfy the bound (\ref{cvisBound}).
In fact, this bound can never be satisfied for $Z=0$ since in this case the leading decay channel of $\phi$
into visible sector fields would be induced by loop-suppressed couplings to gauge bosons \cite{DRinLVS,WinoDMDR}.

The second consequence of $\Delta N_{\rm eff}\lesssim 0.96$ at $2\sigma$ level
is a lower bound on the reheating temperature of order $T_{\rm rh}\gtrsim \mc{O}(1)$ GeV
which disfavours the branching scenario with Bino-like DM.
In fact, using (\ref{Mgaugino}), the expression (\ref{Trh2}) for $T_{\rm rh}$ can be rewritten
in terms of $\Delta N_{\rm eff}$ and $M_{1/2}$ as:
\be
T_{\rm rh}\simeq \frac{\kappa}{\sqrt{\Delta N_{\rm eff}}}\left(\frac{68.5}{g_*}\right)^{1/4}
\left(\frac{M_{1/2}}{1\,{\rm TeV}}\right)^{9/8}\,1.19\,{\rm GeV}\,.
\label{Trh3}
\ee
Combining DR bounds with Fermi data, one obtains the same constraints on the $(\Delta N_{\rm eff},m_{\rm DM})$-plane
as those depicted in Fig. \ref{Fig2} with the further observation that thermal DM scenarios cannot be realised
in sequestered LVS models even for $m_{\rm DM}\gtrsim 40$ GeV. This is because thermal DM can be allowed only for $\Delta N_{\rm eff} \lesssim 0.14$
which correlates with $Z\gtrsim 4.7$. However a large value of $Z$ in (\ref{KHiggs}) would induce a large $\mu$-term
and thus heavy Higgsinos. In this case, the LSP would therefore be Bino-like, resulting in thermal DM overproduction.

Thus we found that \emph{in sequestered LVS models, DR constraints combined with Fermi data prefer non-thermal Higgsino-like DM.}
Moreover, $\Delta N_{\rm eff}$ and $M_{1/2}$ set a lower bound for the DM mass since eq. (\ref{B2}) becomes:
\be
m_{\rm DM} \gtrsim \Delta N_{\rm eff}
\sqrt{\frac{g_*}{68.5}} \left(\frac{1 \,{\rm TeV}}{M_{1/2}}\right)^{9/4}
\,\frac{230\,{\rm GeV}}{\kappa^2}\,.
\label{B4b}
\ee
Given that the right-hand side of the inequality (\ref{B4b}) is a straight line in $\Delta N_{\rm eff}$
whose slope is set by $M_{1/2}^{-9/4}$, this lower bound becomes weaker from larger values of $M_{1/2}$ (see Fig. \ref{Fig3}).
In particular, the central value $\Delta N_{\rm eff}\simeq 0.5$ would give $m_{\rm DM} \gtrsim \mc{O}(110)$ GeV for $g_*=68.5$.

Let us finally point out that if Higgsinos annihilate too efficiently to satisfy the DM relic density,
the DM candidate can be either a well-tempered Bino/Higgsino
or a multi-component axion-Higgsino. As explained in Appendix \ref{App}, sequestered LVS models,
in addition to the volume axion that is the source of DR, feature also an open string mode which behaves as the QCD axion \cite{WinoDMDR}.
The coherent oscillations of this QCD axion can give a relevant contribution to the DM abundance
depending on the value of its decay constant $f_a$ (see Appendix \ref{App}).

\section{Conclusions}
\label{Concl}

In this paper, we have explored connections between DM and DR when they are both sourced by the decay of a string modulus.
We have shown that this common origin is a generic situation in string models with perturbative moduli stabilisation
where the axions remain light because of their shift symmetry. Moreover, bulk closed string axions
do not get eaten up by anomalous $U(1)$s, resulting in the generic production of axionic DR from the decay of the lightest modulus
$\phi$ which drives reheating.

By combining present cosmological bounds on $\Delta N_{\rm eff}$ \cite{Planck} with Fermi data from dwarf spheroidal galaxies \cite{fermi},
we managed to derive interesting constraints on the nature of DM, its mass and production mechanism.
In fact, we found that some regions of the $(\Delta N_{\rm eff}, m_{\rm DM})$-plane are observationally disfavoured
as a function of the modulus mass $m_\phi$ which in string compactifications is tied to the scale of the soft terms.
Most of allowed region of this parameter space corresponds to non-thermal annihilation scenarios with Higgsino-like DM,
while thermal DM can be accommodated only if $N_{\rm eff}$ tends to its SM value.

In the context of sequestered type IIB LVS model \cite{sequestering}, where a modulus mass of order
$m_\phi \simeq 5\cdot 10^6$ GeV generates TeV-scale soft terms, DR overproduction
can be avoided only in the presence of a Giudice-Masiero coupling in the K\"ahler potential.
Moreover, the reheating temperature is bounded from below, $T_{\rm rh}\gtrsim \mc{O}(1)$ GeV,
implying that non-thermal branching scenarios with Bino-like DM are ruled out.
Also thermal scenarios are disfavoured since $\Delta N_{\rm eff}$ close to zero correlates with Bino-like LSP
that leads to DM overproduction. In fact, a very small $\Delta N_{\rm eff}$ can be obtained only
by taking large values of $Z$ that would induce a large $\mu$-term contribution to the Higgsino mass.
Hence DM has to be produced non-thermally as in the annihilation scenario and
it is preferentially Higgsino-type (possibly either well-tempered Higgsino/Bino or multi-component axion-Higgsino).
Finally, we derived a lower bound on the DM mass in terms of $\Delta N_{\rm eff}$ and the unified gaugino mass $M_{1/2}$,
showing that the allowed DM masses can be probed at the LHC, e.g. via Vector Boson Fusion~\cite{Delannoy:2013ata}.

\section{Acknowledgements}

We would like to thank Scott Watson for helpful discussions and Savvas Koushiappas for providing the data from the analysis in \cite{fermi}. This work is supported in part by DOE Grant No. DE-SC0010813 (B.D.) and NASA Astrophysics Theory Grant NNH12ZDA001N (K.S.).

\appendix

\section{Moduli stabilisation, anomalous $U(1)$s and light axions}
\label{App0}

In string compactifications, 4D axions $a_i$, with $i=1,...,N$, emerge as the imaginary components of complex scalar fields $T_i$
whose real parts $\tau_i$ parameterise the size of the extra dimensions \cite{Cicoli:2012sz}.
The number of these moduli fields is counted by topological properties of Calabi-Yau (CY) three-folds,
and generically turns out to be very large: $N\sim \mc{O}(100)$.
Moreover, these axions come along with shift symmetries $a_i \to a_i + {\rm const}$ that are exact in perturbation theory
and are broken only by non-perturbative effects. Hence, the axions can develop a potential only at non-perturbative level.
On the contrary, the geometric moduli $\tau_i$ are not protected by any symmetry,
and so can become massive at both perturbative and non-perturbative level. There are therefore two interesting situations \cite{Nflation}:
\begin{itemize}
\item If the moduli $\tau_i$ are fixed by non-perturbative corrections to the effective action,
their mass is as large as that of the corresponding axions $a_i$ since both fields are stabilised by the same effects.
In this case, the axions become too heavy to play any significant r\^ole since they would need to be heavier than $\mc{O}(50)$ TeV
in order to avoid any cosmological moduli problem (CMP). Notice that the moduli masses are generically of order the gravitino mass
$m_{3/2}$. Hence, if one assumes a solution of the CMP due to a non-standard cosmological evolution of our Universe,
and tries to lower the axion mass to values of $\mc{O}({\rm meV})$ relevant for phenomenology, one would still face the problem of
a very low supersymmetry breaking scale.

\item On the other hand, if the moduli $\tau_i$ are stabilised by perturbative effects, they become heavier than
the corresponding axions $a_i$ whose directions are lifted only at subleading order by tiny non-perturbative effects.
In this case, the mass of the moduli $\tau_i$ can be larger than $\mc{O}(50)$ TeV, avoiding any CMP,
whereas the axions $a_i$ can be very light. One of these $a_i$ fields can behave as the QCD axion
only if stringy instantons or non-perturbative contributions from gaugino condensation give a mass to this field
which is smaller than the one generated by standard QCD instantons.
\end{itemize}

Some of the light axions which emerge in models with perturbative moduli stabilisation,
can still disappear from the low-energy theory by getting eaten up by anomalous $U(1)$ gauge bosons
in the process of anomaly cancellation. More in detail, each anomalous $U(1)$ factor comes along
with a D-term potential of the form \cite{Jockers:2005zy}:
\be
V_D \simeq g^2 \left(\sum_\alpha q_\alpha |C_\alpha|^2-\xi\right)^2\,,
\ee
where $C_\alpha$ are open string matter fields with $U(1)$-charges $q_\alpha$,
and $\xi$ is a moduli-dependent Fayet-Iliopoulos (FI) term. In fact, $\xi$ can be expressed
in terms of the K\"ahler potential $K$ of the 4D $\mc{N}=1$ effective theory as \cite{Jockers:2005zy}:
\be
\xi= -\tilde{q}_i \frac{\partial K}{\partial T_i}M_P^2\,,
\ee
where $\tilde{q}_i$ are the $U(1)$-charges of the $T$-moduli.
Notice that, besides closed string axions $a_i$, one can have also open string axions $\psi_\alpha$
which emerge as the phases of the matter fields $C_\alpha$, and enjoy a shift symmetry
of the form $\psi_\alpha \to \psi_\alpha + q_\alpha$. From now on, for simplicity,
we shall focus on the case with just one matter field $C$ and one modulus $\tau$.

D-term stabilisation fixes a combination of $|C|$ and $\tau$ corresponding
to the combination of open and closed string axions eaten up by the anomalous $U(1)$ \cite{Choi:2006bh}
which acquires a St\"uckelberg mass of the form \cite{ArkaniHamed:1998nu}:
\be
M_{U(1)}^2 \simeq g^2 \left[\left(f_a^{\rm open}\right)^2+\left(f_a^{\rm closed}\right)^2\right]\,.
\label{MU1}
\ee
The open string axion decay constant $f_a^{\rm open}$ is set by $|C|$
which, in turn, is fixed in terms of $\xi$ by imposing $V_D=0$:
\be
\left(f_a^{\rm open}\right)^2= |C|^2 =\frac{\xi}{q} = -\frac{\tilde{q}}{q} \frac{\partial K}{\partial T}M_P^2\,.
\label{fopen}
\ee
On the contrary, the closed string axion decay constant $f_a^{\rm closed}$ is determined from canonical normalisation \cite{Cicoli:2012sz}:
\be
\left(f_a^{\rm closed}\right)^2 = \frac{\partial^2 K}{\partial T^2}M_P^2\,.
\label{fclosed}
\ee
If in (\ref{MU1}), $f_a^{\rm open} \gg f_a^{\rm closed}$, the combination of axions eaten up by the anomalous $U(1)$ is mostly $\psi$,
whereas if $f_a^{\rm open} \ll f_a^{\rm closed}$, the axion removed from the low-energy spectrum is $a$.

In type IIB models, non-Abelian gauge theories with chiral matter can live on either
magnetised D7-branes wrapped around four-cycles in the geometric regime or fractional D3-branes
at singularities. In the geometric case, the K\"ahler potential is $K=-3\ln(T+\overline{T})$,
and so the open string axions are eaten up since:
\be
\left(f_a^{\rm open}\right)^2 = \left(\frac{3\tilde{q}}{2 q}\right)\frac{M_P^2}{\tau}\gg
\left(f_a^{\rm closed}\right)^2 = \frac{3M_P^2}{4\tau^2}
\quad\text{for}\quad\tau\gg 1\,. \nn
\ee
Hence we showed the following important result: \emph{the presence of light closed string axions is a model-independent prediction
of any compactification where some moduli are stabilised perturbatively}.
As can be seen from (\ref{MU1}), the $U(1)$ mass is of order the Kaluza-Klein
scale $M_{\rm KK}$: $M_{U(1),{\rm geom}}\simeq M_{\rm KK} \simeq M_P/\vo^{2/3}$, since $g^2\simeq 1/\tau$
and the CY volume is $\vo\simeq \tau^{3/2}$. In the second case, $K$ has a different functional dependence
on the modulus $\tau_{\rm sing}$ which resolves the singularity where D3-branes are localised:
$K= (T_{\rm sing}+\overline{T}_{\rm sing})^2/\vo$. Thus in this case the axions eaten up are the closed string ones
since (for $q<0$ and $\tau_{\rm sing}\ll 1$):
\be
\left(f_a^{\rm open}\right)^2 = \left(\frac{4\tilde{q}}{|q|}\right)\frac{\tau_{\rm sing}M_P^2}{\vo}\ll
\left(f_a^{\rm closed}\right)^2 = \frac{2 M_P^2}{\vo}\,.
\label{faopen}
\ee
The anomalous $U(1)$ acquires a mass from (\ref{MU1}) of order the string scale $M_s$:
$M_{U(1),{\rm sing}} \simeq M_s \simeq M_P/\vo^{1/2}$ since in the singular regime $g^2$
does not depend on the geometric moduli (it is set by the string coupling).

\section{Axions in sequestered LVS models}
\label{App}

Let us now study the r\^ole played by different open and closed string axions
in type IIB LVS models where the visible sector is sequestered from supersymmetry breaking \cite{sequestering}.
These constructions are particularly interesting since the moduli mass spectrum and couplings can be computed explicitly,
allowing a detailed analysis of the post-inflationary cosmological evolution.
Moreover, these models are characterised by a high gravitino mass that allows for successful inflationary
model building, and guarantees the absence of any gravitino and cosmological moduli problem, together with
low-energy supersymmetry because of sequestering.

The simplest version of these models involves a CY with volume form:\footnote{See \cite{ExplicitLVS,Cicoli:2013cha} for explicit realisations
of sequestered models.}
\be
\vo\,= \tau_b^{3/2}-\tau_{\rm np}^{3/2} - \tau_{\rm vs}^{3/2}\,,
\ee
where $\tau_b$ is a `big' cycle controlling the size of the extra dimensions,
$\tau_{\rm np}$ is a blow-up mode supporting non-perturbative effects and $\tau_{\rm vs}$
is the visible sector cycle which is stabilised supersymmetrically by D-terms at zero size:
$\tau_{\rm vs}\to 0$. The visible sector lives on spacetime filling D3-branes localised at
the singularity obtained by shrinking $\tau_{\rm vs}$. Supersymmetry is broken
by the F-terms of the bulk moduli which are fixed by the following effects \cite{LVS}:
\begin{itemize}
\item $\tau_{\rm np}$ is fixed by non-perturbative effects, and so
the corresponding axion $a_{\rm np}$ develops a mass of the same order of magnitude:
\be
m_{\tau_{\rm np}}\simeq m_{a_{\rm np}}\simeq m_{3/2}\ln\left(\frac{M_P}{m_{3/2}}\right)\,.
\label{manp}
\ee

\item $\tau_b$ is fixed because of perturbative $\alpha'$ effects, and so the corresponding axion
$a_b$ is lighter due to its shift symmetry. In fact, the mass of $\tau_b$ is of order:
\be
m_{\tau_b}\simeq m_{3/2}\sqrt{\frac{m_{3/2}}{M_P}}\ll m_{3/2}< m_{\tau_{\rm np}}\,,
\label{mtb}
\ee
whereas the mass of the bulk axion $a_b$ generated by $T_b$-dependent non-perturbative effects, scales as:
\be
m_{a_b} \simeq M_P\, e^{-2\pi \left(\frac{M_P}{m_{3/2}}\right)^{2/3}}\ll m_{\tau_b}\,.
\label{mab}
\ee
\end{itemize}
In these sequestered models, the soft-terms generated by gravity mediation are suppressed with respect to the gravitino mass.
In fact, the unified gaugino masses behave as \cite{sequestering}:
\be
M_{1/2}\simeq m_{3/2}\left(\frac{m_{3/2}}{M_P}\right) \ll m_{\tau_b}\,,
\label{M12}
\ee
while the unified scalar masses scale as:
\be
m_0 \simeq m_{3/2}\left(\frac{m_{3/2}}{M_P}\right)^\alpha\,,
\ee
where the parameter $\alpha$ can be either $1/2$ or $1$ depending on the moduli-dependence of the K\"ahler metric
for matter fields~\cite{sequestering}. If the physical Yukawa couplings $Y_{abc}$ are required to be independent
of the CY volume $\vo$ at leading order, one is in the local limit
where $\alpha=1/2$ and $m_0\simeq m_{\tau_b}$, whereas if the $Y_{abc}$ do not depend on $\vo$ also at subleading order,
one is in the ultra-local limit where $\alpha=1$ and $m_0\simeq M_{1/2}$.
Due to lack of an explicit computation to determine the K\"ahler metric for matter fields,
both of these options seem to be physically possible.

From (\ref{M12}), it can be easily seen that TeV-scale gauginos can be obtained for $m_{3/2} \simeq \mc{O}(5\cdot 10^{10})$ GeV
which from (\ref{manp}) implies $m_{\tau_{\rm np}}\simeq m_{a_{\rm np}}\simeq\mc{O}(10^{12})$ GeV,
from (\ref{mtb}) $m_{\tau_b}\simeq \mc{O}(5\cdot 10^6)$ GeV and from (\ref{mab}) $m_{a_b}\sim 0$.
The volume mode $\tau_b$ is therefore the lightest modulus whose decay reheats the visible sector degrees of freedom.
Given that $50\, {\rm TeV}\ll m_{\tau_b}\ll m_{3/2}$, both the CMP and the gravitino problem are automatically absent
in these constructions.

Let us now analyse the behaviour of closed and opened string axions in sequestered LVS models:
\ben
\item \emph{Closed string axions}: as we have seen $a_{\rm np}$ is very heavy, and so decouples from low-energy physics.
On the other hand, the volume axion $a_b$ is almost massless. Moreover, as we have shown in the previous section,
it does not get eaten up by any anomalous $U(1)$. Thus we conclude that \emph{the presence of ultra-light bulk axions
is a model-independent prediction of LVS models}. Due to its geometric separation from
the visible sector localised on branes at singularities, $a_b$ does not couple to QCD, and so
cannot play the r\^ole of the QCD axion \cite{Cicoli:2012sz}. However it can behave as DR produced
from the decay of $\tau_b$. Therefore we also found that \emph{DR is a generic prediction of LVS compactifications}.
We finally mention that, for any del Pezzo singularity, all closed string axions associated to cycles which collapse
to zero size like $\tau_{\rm vs}$, get eaten up by anomalous $U(1)$s.

\item \emph{Open string axions}: these fields survive in the low-energy theory only if
they are localised at singularities since in the previous section we have shown that
in geometric regime they tend to get eaten up by anomalous $U(1)$s.
Focusing on the singular case and recalling that $m_{3/2}\simeq M_P/\vo$, (\ref{faopen}) gives:
\be
f_a^{\rm open}\simeq |C| \simeq M_s \sqrt{\tau_{\rm sing}}\,,
\label{faopen2}
\ee
where $M_s\simeq \sqrt{m_{3/2}M_P}\simeq \mc{O}(5\cdot 10^{14})$ GeV.
Thus the exact value of the axion decay constant depends on the stabilisation of the matter
field $|C|$ which develops a potential due to SUSY-breaking effects and $\tau_{\rm sing}$-dependent
terms which break the no-scale structure once they are written in terms of $|C|$ by using (\ref{faopen2}) \cite{Cicoli:2013cha}.
If $|C|$ acquires a positive mass-squared from supersymmetry breaking, $|C|=0$,
and so $\tau_{\rm sing} = 0$, ensuring that this cycle is indeed collapsed to zero size.
However this case does not lead to a viable QCD axion unless $|C|$ develops a non-vanishing VEV
due to RG running below the string scale.
On the other hand, if the soft scalar mass of $|C|$ is tachyonic,
$|C|$ gets a non-zero VEV which in the local case is of order \cite{Cicoli:2013cha}:
\be
f_a^{\rm open} \simeq |C| \simeq M_s \sqrt{\frac{m_{3/2}}{M_P}}
\simeq \mc{O}(5\cdot 10^{10})\,\text{GeV}\,, \nn
\ee
while in the ultra-local case looks like:
\be
f_a^{\rm open} \simeq |C| \simeq M_s \left(\frac{m_{3/2}}{M_P}\right)^{3/2}
\simeq \mc{O}(1)\,\text{TeV}\,. \nn
\ee
In both situations, $\tau_{\rm sing}$ is still in the singular regime, since
$\tau_{\rm sing} \simeq m_{3/2}/M_P \ll 1$ in the local case,
whereas $\tau_{\rm sing}\simeq (m_{3/2}/M_P)^3\ll 1$ in the ultra-local case.
However, the phase $\psi$ of $C$ can be the QCD axion only for the local case
which gives a decay constant inside the phenomenologically allowed window:
$10^9\,\text{GeV}\lesssim f_{\rm QCD}\lesssim10^{12}\,\text{GeV}$.
On the other hand, the ultra-local case seems to be ruled out.
The only way-out for this situation would be to focus on the case where the VEV of $|C|$
is zero at the string scale but becomes non-vanishing at a lower scale because of RG running
effects. We finally note that $C$ has to be a SM singlet since otherwise its large VEV
would break any SM symmetry at a high scale.
\een

\end{document}